\documentclass[conference]{IEEEtran}
\IEEEoverridecommandlockouts
\usepackage{cite}
\usepackage{amsmath,amssymb,amsfonts}
\usepackage{algorithmic}
\usepackage{graphicx}
\usepackage{textcomp}
\usepackage{xcolor}
\def\BibTeX{{\rm B\kern-.05em{\sc i\kern-.025em b}\kern-.08em
    T\kern-.1667em\lower.7ex\hbox{E}\kern-.125emX}}
    
\usepackage[absolute]{textpos}

\begin{document}

\begin{textblock}{10}(2.8,0.5)
\noindent{\footnotesize \normalfont This is the authors' extended version. The authoritative version will appear in the proceedings of ACM/IEEE SEC'20.}
\end{textblock}

\title{Third ArchEdge Workshop: Exploring the Design Space of Efficient Deep Neural Networks}


\author{
    \IEEEauthorblockN{\emph{Invited paper}\IEEEauthorrefmark{4}\\~\\Fuxun Yu\IEEEauthorrefmark{1}, Dimitrios Stamoulis\IEEEauthorrefmark{2}, Di Wang\IEEEauthorrefmark{2}, Dimitrios Lymberopoulos\IEEEauthorrefmark{2}, Xiang Chen\IEEEauthorrefmark{1}}
    \IEEEauthorblockA{\IEEEauthorrefmark{1}George Mason University, \IEEEauthorrefmark{2}Microsoft}
    \thanks{\IEEEauthorrefmark{4}Presented at the Third Workshop on Computing Architecture for Edge Computing (ArchEdge), co-located with the fifth ACM/IEEE Symposium on Edge Computing (SEC), November 11-13, 2020. Email: fyu2@gmu.edu}
    }

\maketitle

\begin{abstract}

This paper gives an overview of our ongoing work on the design space exploration of efficient deep neural networks (DNNs). Specifically, we cover two aspects: (1) static architecture design efficiency and (2) dynamic model execution efficiency. For static architecture design, different from existing ``end-to-end'' hardware modeling assumptions, we conduct ``full-stack'' profiling at the GPU core level to identify better accuracy-latency trade-offs for DNN designs. For dynamic model execution, different from prior work that tackles model redundancy at the DNN-channels level, we explore a new dimension of DNN \textit{feature map redundancy} to be dynamically traversed at runtime. Last, we highlight several open questions that are poised to draw research attention in the next few years.

\end{abstract}

\section{Introduction}

This paper summarizes our latest explorations in the design space of efficient DNNs. Specifically, we cover two complementary aspects: (a) static architecture design efficiency and (b) dynamic model execution efficiency.

\subsection{Static Architecture Design Efficiency}

Recent AutoML techniques, such as Neural Architecture Search (NAS), aim to automate the design of DNNs~\cite{wen2020neural, cheng2020nasgem, zoph2016neural}. Due to the considerable search cost required to traverse the DNN design space, early approaches can take thousands of GPU hours to select the final model candidates~\cite{gpu_hours}. To this end, in the context of hardware-constrained DNNs, previous work has significantly improved the  search efficiency thanks to hardware performance models (e.g., DNN FLOPs, energy consumption, latency, etc.) which allow the AutoML algorithm to efficiently traverse the design space in a hardware-aware manner~\cite{ mnasnet, effnet, stamoulis2020single}, by quickly discarding DNNs that violate the hardware constraints of the target platform. 

Hardware-aware AutoML algorithms resort to certain ineffective performance interpretations with respect to the underlying hardware. On the one hand, early AutoML methods~\cite{darts} use FLOPs as a general performance indicator, yet recent works demonstrate a mismatch between FLOPs and hardware metrics~\cite{dong2018dpp, marculescu2018hardware, cai2017neuralpower, yang2017designing}. Nevertheless,  this mismatch has been discussed mainly through empirical results and is not comprehensively analyzed. On the other hand, while recent methods replace FLOPs with predictive models (e.g., latency, power consumption), they rely on ``end-to-end'' profiling, which is either limited to discrete design choices (e.g., 50\%, 100\% channels)~\cite{stamoulis2018hyperpower} or follows a look-up table-based manner~\cite{stamoulis2020single}. 

To this end, we present a comprehensive ``full-stack'' profiling analysis that dives into individual GPU cores/threads to examine the intrinsic mechanisms of DNN execution~\cite{our_mlsys}. As a key contribution, we shed light into the ``GPU tail'' effect as root cause of FLOPs-latency mismatch and GPU under-utilization. Based on our findings, we revisit the DNN design configuration choices of state-of-the-art AutoML methodologies to eliminate the tail effect,  enabling larger, more accurate DNN designs \textbf{at no latency cost}. Hence, our method \textit{concretely improves accuracy-latency trade-offs}, such as 27\% latency and 4\% accuracy improvements on top of SOTA DNN pruning and NAS methods. Moreover, we extend our profiling finding across different GPU configurations.

\textbf{Discussion - Future work}: while our investigation is employed as a fine-tuning (local search) step on top of SOTA designs, our findings can be flexibly incorporated into other AutoML methods. That is, a direction for future work is to revisit the predictive-models of existing single- and multi-path NAS works~\cite{single_path, cai2018proxylessnas} to further improve the accuracy-latency trade-offs by traversing the design space in a ``tail effect''-aware fashion. Moreover, our methodology focuses on eliminating ``tail effects'' at the DNN design level, but improvements from alleviating GPU under-utilization can be realized at other design levels, as shown by novel scheduling- and computational flow-level explorations~\cite{ding2020ios, yu2020dc}. 

Next, we hope that our findings could inspire researchers to revisit design-space assumptions, by allowing to identify hardware-optimal DNN candidates while eliminating sub-optimal ones. For example, our channel-level analysis  reveals a \textit{discrete} set of DNN channel configurations~\cite{our_mlsys} with optimal GPU utilization, which could potentially reduce the number of candidates by $10\times$ as opposed to traversing a \textit{continuous} channel-number space, e.g., on top of existing channel-pruning methods~\cite{amc, chin2020towards}. Last, an interesting direction would be to investigate the severity of similar under-utilization beyond GPUs, especially in the context of hardware accelerators and co-design NAS methodologies~\cite{zhang2020dna}.

\subsection{Dynamic Model Execution Efficiency}

Dynamic execution methods aim at selecting between ``switchable'' DNN components at runtime~\cite{xu2020directx, chen2020dynamic, dynamic1, dynamic2, xu2019reform}. The key insight behind these works is to improve the  overall model efficiency by adaptively selecting and executing (a subset of) the model based on the input characteristics~\cite{stamoulis2018designing}. In our work, we extend this intuition across a new \textit{model redundancy} dimension, namely dynamic feature map redundancy~\cite{our_date}. 
Specifically, we show that feature redundancy exists at the spatial dimensions of DNN convolutions, which allows us to formulate a dynamic pruning methodology 
in \textbf{both channel- and spatial-wise dimensions}. Our proposed method can greatly reduce the model computation with up to 54.5\% FLOPs reduction and
negligible accuracy drop on various image-classification DNNs.

\textbf{Discussion - Future work}: Drawing inspiration from our analysis on the FLOPs-latency mismatch, we highlight that when implemented naively, merely pruning convolution weights at the spatial level does not translate to latency savings. To this end, we postulate that advances in sparse DNN operators will be essential to support \textit{dynamic-sparse} execution, as recently shown with CUDA implementations for dynamic convolutions~\cite{verelst2020dynamic}.

\section{Conclusion} 
In this paper, we summarize a set of novel efficiency optimization angles for DNN design in both static architecture design and dynamic model execution. New potential advantages can be attained by integrating the proposed new perspectives to current optimization methods.

\vspace{-8pt}


~\\
\bibliography{main}
\bibliographystyle{IEEEtran}

\end{document}